\documentclass[aps,pra,showpacs,superscriptaddress,preprint]{revtex4}
\usepackage{amsmath}
\usepackage{epsf}
\usepackage{epsfig}

\begin{document}

\title{Coherent States for PT-/Non-PT-Symmetric and Non-Hermitian Morse
Potential via Path Integral Method}
\author{\small Nalan Kandirmaz}
\email[E-mail: ]{nkandirmaz@mersin.edu.tr}\affiliation{Department of
Physics, Mersin University, Mersin,Turkey}
\author{\small Ramazan Sever}
\email[E-mail: ]{sever@metu.edu.tr}\affiliation{Department of
Physics, Middle East Technical  University, Ankara,Turkey}
\date{\today}

\begin{abstract}
We discuss the coherent states for PT-/non-PT-Symmetric and
non-Hermitian generalized Morse Potential obtained by using path
integral formalism over the holomorphic coordinates. We transform
the action of generalized Morse potential into two harmonic
oscillators with a new parametric time to establish the parametric
time coherent states. We calculate the energy eigenvalues and the
corresponding wave functions in parabolic coordinates.

\end{abstract}


\maketitle

\address{$^a$ Mersin University, Department of Physics,Mersin\\

$^b$ Middle East Technical University, Department of Physics,Ankara}

\section{Introduction}

PT-symmetric quantum mechanics has been widely investigated
recently. In standard formalism of quantum mechanics, Hamiltonian
that defines the dynamics and symmetries of system is Hermitian.
In the PT-symmetric case, Hamiltonian has real spectra although it
is not Hermitian. Bender and his co-workers studied PT-symmetric
Hamiltonians which have real eigenvalues \cite{Bender}. Following
these works, many authors studied PT-symmetric and non-Hermitian
Hamiltonians having real and/or complex eigenvalues
\cite{Ahmed,Bagchi,Znojil} by applying the analytic methods and
numerical techniques. In this direction, several methods have been
developed to calculate energy eigenvalues and the wave functions
\cite{Berkdemir2}. One of the methods providing exact solutions is
Path integral method . It is also employed to find the transition
amplitudes among the configuration space eigenstates
\cite{Feynman}. The energy spectra and eigenfunctions of
PT-/non-PT- and non-Hermitian Morse potential is obteined
\cite{Sever1}.

Morse potential which has an important role in molecular physics is
used for the description of the interaction between the atoms in
diatomic molecules \cite{Morse,Unal2, Berkdemir1}. Bound-state
energy spectrum and the wave functions of Morse potentials is
obtained by Duru using path integral method \cite{Duru1}. Later Unal
derived parametric time coherent states using path integrals over
holomorphic coordinates \cite{Unal2}. Coherent states which are
firstly constructed by Schr\"{o}dinger are defined as eigenstates of
lowering operators for harmonic oscillators \cite{Schrodinger}. They
are minimum uncertainty states and do not disperse during the
evolution of the system \cite{Unal1,Unal2,Unal4}. Both PT-symmetric
and non-Hermitian generalized Morse potential have been studied
exactly \cite{Sever1,Sever2}. In the present work coherent states of
PT-/non-PT- and non-Hermitian Morse potential is discussed by using
Path Integral method. We obtain the classical dynamics in terms of
the holomorphic coordinates because of that holomorphic coordinates
are the classical analogue of the rising and lowering operators of
the harmonic oscillators. Energy eigenvalues and the corresponding
wave functions are obtained by transforming the problem into the two
oscillators in the parametric time and quantizing these oscillators
using the path integration over the holomorphic coordinates. The
path integral technique can be applied to any potential if it is
transformed into the Harmonic oscillator potential.

 The organization of the paper is as follows: In Sec. 2, we introduce coherent states of the
generalized Morse potential. We obtain the energy eigenvalues and
wave functions in parabolic coordinates. In Secs. 3 and 4, we get
the coherent states, energy eigenvalues and corresponding
eigenfunctions of the $PT$-symmetric and non-$PT$-symmetric
non-Hermitian forms of the generalized Morse potential. Finally,
Results are discussed in Sec. 5.

\section{Generalized Morse Potential}

We consider a particle with mass m moving in generalized Morse
potential
\begin{equation}
V(x)=V_{1}e^{-2\alpha x}-V_{2}e^{-\alpha x}
\end{equation}
where ${V}_{1}$, $V_{2}$ are constants and $\alpha$ is the width
of the potential. The action of the one dimensional generalized
Morse potential is
\begin{equation}
A=\int dt\left[ p_{x}\frac{dx}{dt}-\left( \frac{p_{x}^{2}}{2m}%
+V_{1}e^{-2\alpha x}-V_{2}e^{-\alpha x}\right) \right].
\end{equation}
Let us define a new coordinate u(t)
\begin{equation}
x=-2\alpha \ln u
\end{equation}
and canonical conjugate momentum
\begin{equation}
p_{x}=-\frac{\alpha u}{2}p_{u}.
\end{equation}
After this transformation, action takes{\normalsize
\begin{equation}
A=\int dt\left[ p_{u}\overset{\cdot }{u}-\left( \frac{\alpha ^{2}u^{2}}{4}%
\frac{p_{u}^{2}}{2m}+V_{1}u^{4}-V_{2}u^{2}\right) \right].
\end{equation}
In order to annihilate the factorization u}$^{2}$ in kinetic
energy term, we define a new  time parameter as
\begin{equation}
\frac{dt}{d\tau }=\frac{1}{u^{2}}
\end{equation}
Substitution of this new time parameter into Eq. (5) gives us
\begin{equation}
A=\int d\tau \left[ p_{u}\frac{du}{d\tau }-\left( \frac{\alpha ^{2}}{4}\frac{%
p_{u}^{2}}{2m}+V_{1}u^{2}-V_{2}\right) +\left( -p_{0}\right) \left( \frac{dt%
}{d\tau }-\frac{1}{u^{2}}\right) \right].
\end{equation}
Here an additional Lagrange multiplier $p_{0}$ is used due to the presence of a new parametric time. Adding a dummy momentum $p_{\phi }$ by $%
\frac{d\phi }{d\tau }$ \ another Lagrange multiplier. Action
becomes
\begin{equation*}
A=\int d\tau \left[ p_{u}\frac{du}{d\tau }-\frac{1}{2M}\left(
p_{u}^{2}+2MV_{1}u^{2}-2MV_{2}+\frac{p_{\phi }^{2}}{u^{2}}\right)
\right.
\end{equation*}
\begin{equation}
+\left. \left( -p_{0}\right) \frac{dt}{d\tau }+\frac{d\phi }{d\tau
}\left( p_{\phi }-\sqrt{-2Mp_{0}}\right) \right],
\end{equation}
{\normalsize where }$M=\frac{4m}{\alpha ^{2}}$. Let us define
{\normalsize $\omega =\sqrt{\frac{2V_{1}}{M}}$ as the frequency of
the two oscillators. So that Eq. (8) can be rewritten as
\begin{equation*}
A=\int d\tau \left[ p_{u}\frac{du}{d\tau }-\frac{1}{2M}\left( p_{u}^{2}+%
\frac{1}{2}M^{2}\omega ^{2}u^{2}+\frac{p_{\phi
}^{2}}{u^{2}}-2MV_{2}\right) \right.
\end{equation*}
\begin{equation}
\left. +\left( -p_{0}\right) \frac{dt}{d\tau }+\frac{d\phi }{d\tau
}\left( p_{\phi }-\sqrt{-2Mp_{0}}\right) \right]
\end{equation}
We represent the position vector of the particle in the two
dimensional polar coordinates as
\begin{equation}
\overrightarrow{\mathbf{u}}\mathbf{=}\left( u_{1},u_{2}\right) =\sqrt{u}%
\left( \cos \phi ,\sin \phi \right).
\end{equation}
Action in terms of $\left(u_{1},u_{2}\right)$ becomes
\begin{equation*}
A=\int d\tau \left[ p_{u_{1}}\frac{du_{1}}{d\tau }+p_{u_{2}}\frac{du_{2}}{%
d\tau }-\omega \left[ \frac{p_{u_{1}}^{2}+p_{u_{2}}^{2}}{2M\omega }+\frac{1}{%
2}M\omega (u_{1}^{2}+u_{2}^{2})-\frac{V_{2}}{\omega }\right]
\right.
\end{equation*}
\begin{equation}
+\left. \left( -p_{0}\right) \frac{dt}{d\tau }-\sqrt{-2Mp_{0}}\frac{d\phi }{%
d\tau }\right]
\end{equation}
{\normalsize Here one can define $A^{\prime }$ as an action of two
oscillators system with mass $M$ frequency $\omega $ and energy
$V_{2}$. Thus we have
\begin{equation}
A^{\prime }=\int d\tau \left\{ p_{u_{1}}\frac{du_{1}}{d\tau }+p_{u_{2}}\frac{%
du_{2}}{d\tau }-\omega \left[ \frac{p_{u_{1}}^{2}+p_{u_{2}}^{2}}{2M\omega }+%
\frac{1}{2}M\omega (u_{1}^{2}+u_{2}^{2})-\frac{V_{2}}{\omega
}\right] \right\}.
\end{equation}
The term
$\left[\frac{p_{u_{1}}^{2}+p_{u_{2}}^{2}}{2M}+\frac{1}{2}M\omega
^{2}(u_{1}^{2}+u_{2}^{2})\right] -V_{2}$ corresponds to the
Hamiltonian of two oscillators.

 The holomorphic coordinates are
defined as
\begin{equation}
a=\binom{a_{1}}{a_{2}}=\frac{1}{\sqrt{2}}\binom{\sqrt{m\omega }u_{1}+i\sqrt{%
\frac{1}{M\omega }}p_{u_{1}}}{\sqrt{m\omega }u_{2}+i\sqrt{\frac{1}{M\omega }}%
p_{u_{2}}}
\end{equation}
and also conjugate dynamical variables as
\begin{equation}
a^{\dagger }=\left( a_{1}^{\ast }~~~a_{2}^{\ast }\right) =\frac{1}{\sqrt{2}}%
\left( \sqrt{m\omega }u_{j}-i\sqrt{\frac{1}{M\omega
}}p_{u_{j}}\right) ~,~~~~\ ~j=1,2.
\end{equation}
The action in Eq. (12) is rewritten again in terms of the
holomorphic coordinates
\begin{equation}
A^{\prime }\left(a_{b}^{\dagger },\tau
_{b};a_{a},\tau_{a}\right)=\ V_{2}\left( \tau _{b}-\tau
_{a}\right)+\int\ d\tau \left[ \frac{1}{2i}\left( \frac{
da^{\dagger }}{d\tau }a-a^{\dagger }\frac{da}{d\tau }\right)
-\omega \left( a^{\dagger }a\right) \right].
\end{equation}
We ignore the total derivative $\frac{1}{2}\sum\limits_{j=1}^{2}
\left. \left( p_{u_{j}}u_{j}\right) \right\vert _{\tau _{a}}^{\tau
_{b}}$ in the action, Eq. (15). The kernel of two oscillators with
parametric time $\tau $ in holomorphic coordinates is
\begin{equation}
K^{\prime }(a_{b}^{\dagger },\tau _{b};a_{a},\tau
_{a})=e^{i(V_{2}-
)(\tau _{b}-\tau _{a})}\int \frac{Da^{\dagger }Da}{\left[ 2\pi i\right] ^{2}}%
\exp\left[i\int\limits_{\tau _{a},a_{a}}^{\tau _{b},a_{b}^{\dagger }}d\tau \frac{%
1}{2i}\left( \frac{da^{\dagger }}{d\tau }a-a^{\dagger }\frac{da}{d\tau }%
\right) -\omega \left( a^{\dagger }a\right)\right].
\end{equation}
Here we take $\hbar =1$ and the last term $\omega $
is appeared from the quantum ordering terms between the operators $\widehat{a}%
^{\dagger }$ and $\widehat{a}$. After the integrations over  $a$, $%
a^{\dagger }$ \cite{Unal2} we get
\begin{equation}
K^{\prime }(a_{b}^{\dagger },\tau _{b};a_{a},\tau _{a})=e^{i\left(
V_{2}-\omega \right) \left( \tau _{b}-\tau _{a}\right) }\exp\left[
a_{b}^{\dagger }a_{a}e^{-i\omega \left( \tau _{b}-\tau _{a}\right)
}\right].
\end{equation}
Since the transformation in Eq. (16) has double value, the
physical kernel becomes
\begin{equation*}
K^{\prime }(a_{b}^{\dagger },\tau _{b};a_{a},\tau _{a})=e^{i\left(
V_{2}-\omega \right) \left( \tau _{b}-\tau _{a}\right) }
\end{equation*}
\begin{equation}
\times \left\{ \exp \left[ a_{b}^{\dagger }a_{a}e^{-i\omega \left(
\tau _{b}-\tau _{a}\right) }\right] +\exp \left[ -a_{b}^{\dagger
}a_{a}e^{-i\omega \left( \tau _{b}-\tau _{a}\right)
}\right]\right\}.
\end{equation}
If the exponential term is expanded into power series of }$a_{b}^{\dagger }$ we have%
\begin{equation}
K^{\prime }\left( a_{b}^{\dagger },\tau _{b};a_{a},\tau
_{a}\right) =e^{\tau _{b}-\tau_{a}}\sum\limits_{n_{1,2}=0}^{\infty
}\left[ 1+\left( -1\right)
^{n_{1}+n_{2}}\right] e^{-i\left[ \left( n_{1}+n_{2}+1\right) \omega -V_{2}%
\right] \left( \tau _{b}-\tau _{a}\right)
}\prod\limits_{j=1}^{2}\frac{ \left( a_{jb}^{\ast }a_{ja}\right)
^{n_{j}}}{\Gamma \left( n_{j}+1\right) }
\end{equation}
where }$n_{1}$ and $n_{2}$ are quantum numbers. They can be
expressed in terms of the radial and angular quantum numbers
$n_{r}$ and $m$ of two oscillators as
\begin{equation}
n_{1}=n_{r}+\frac{\left\vert m\right\vert +m}{2}~,\ n_{2}=n_{r}+\frac{%
\left\vert m\right\vert -m}{2}~.
\end{equation}
and let us take $a_{\mp }$ final eigenstates
\begin{equation}
a_{\mp }=\frac{\left( a_{1b}\mp ia_{2b}\right) ^{\ast }}{\sqrt{2}}
\end{equation}
and $\lambda _{\mp }$ initial eigenstates
\begin{equation}
\lambda _{\mp }=\frac{\left( a_{1a}\mp ia_{2a}\right) }{\sqrt{2}}.
\end{equation}
Thus, we rewrite the Kernel in Eq.(19) as
\begin{equation*}
K^{\prime }\left( a_{\mp },\tau ;\lambda _{\mp };\tau _{a}\right)
=\sum\limits_{n_{r}=0}^{\infty }\sum\limits_{m=-\infty }^{\infty
}e^{-2i \left[ \left( n_{r}+\frac{\left\vert m\right\vert
}{2}+\frac{1}{2}\right) \omega -\frac{V_{2}}{2}\right] \left( \tau
_{b}-\tau _{a}\right) }
\end{equation*}
\begin{equation}
\times \left[ 1+\left( -1\right) ^{2\left\vert m\right\vert
}\frac{\left( a_{+}^{\ast }\lambda _{+}\right) ^{n_{1}}\left(
a_{-}^{\ast }\lambda _{-}\right) ^{n_{2}}}{\Gamma \left(
n_{1}+1\right) \Gamma \left( n_{2}+1\right) }\right].
\end{equation}
We suppose that $\phi $  varies as $-\pi <\phi <\pi $ for derive
in Eq. (23). However the generalized Morse potential is not
periodic in this interval. If we let $\phi \longrightarrow 2\pi
\phi /2L$ and taking the limit $L\longrightarrow \infty $\ Eq.
$\left( 23\right) $, it becomes
\begin{equation*}
K^{\prime }\left( a_{\mp },\tau ;\lambda _{\mp };\tau _{a}\right)
=\sum\limits_{n_{r}=0}^{\infty }\int\limits_{-\infty }^{\infty
}dm~e^{-2i \left[ \left( n_{r}+\frac{\left\vert m\right\vert
}{2}+\frac{1}{2}\right) \omega -\frac{V_{2}}{2}\right] \left( \tau
_{b}-\tau _{a}\right) }
\end{equation*}
\begin{equation}
\left[ 1+\left( -1\right) ^{2\left\vert m\right\vert }\frac{\left(
a_{+}^{\ast }\lambda _{+}\right) ^{n_{1}}\left(
a_{-}^{\ast}\lambda _{-}\right) ^{n_{2}}}{\Gamma \left(
n_{1}+1\right) \Gamma \left( n_{2}+1\right) }\right].
\end{equation}
$K^{\prime }$ can be written in terms of oscillator energy
eigenstates $ \left\vert n_{r},m\right\rangle $ as
\begin{equation}
K^{\prime }\left( a_{\mp },\tau ;\lambda _{\mp };\tau _{a}\right)
=\sum\limits_{n_{r}=0}^{\infty }\int\limits_{-\infty }^{\infty
}dm~\left\langle n_{r},m\right\vert \left. a_{\mp }\right\rangle
^{\ast }\left\langle n_{r},m\right\vert U\left( \tau _{b}-\tau
_{a}\right) \left\vert \lambda _{\mp }\right\rangle ,
\end{equation}
\bigskip where $U\left( \tau _{b}-\tau _{a}\right) $ is parametric time
evolution operator between initial coherent states of the two
oscillators. We can denote
\begin{equation}
\left\vert a_{\mp }\right\rangle =e^{\left( \tau _{b}-\tau
_{a}\right) }\sum\limits_{n_{r}=0}^{\infty }\int\limits_{-\infty
}^{\infty }dm\left[ 1+\left( -1\right) ^{\left\vert m\right\vert
}\right] \frac{\left(
a_{+}^{\ast }\right) ^{n_{1}}\left( a_{-}^{\ast }\right) ^{n_{2}}}{\sqrt{%
\Gamma \left( n_{1}+1\right) \Gamma \left( n_{2}+1\right) }}
\end{equation}
and\bigskip
\begin{equation}
U\left( \tau _{b}-\tau _{a}\right) \left\vert \lambda _{\mp
}\right\rangle =\sum\limits_{n_{r}=0}^{\infty
}\int\limits_{-\infty }^{\infty }dme^{-2i \left[ \left(
n_{r}+\frac{\left\vert m\right\vert }{2}+\frac{1}{2}\right) \omega
-\frac{V_{2}}{2}\right] \left( \tau _{b}-\tau _{a}\right) }\left[
1+\left( -1\right) ^{\left\vert m\right\vert }\right] \frac{\left(
\lambda _{+}\right) ^{n_{1}}\left( \lambda _{-}\right)
^{n_{2}}}{\sqrt{\Gamma \left( n_{1}+1\right) \Gamma \left(
n_{2}+1\right) }}.
\end{equation}
To get the coherent states as a function of physical coordinates,
Green's function of the system is obtained as\bigskip
\begin{equation}
G^{\prime }\left( u_{b},t_{b};\lambda _{\mp },t_{a}\right)
=\sum\limits_{n_{r}=0}^{\infty }\int\limits_{-\infty }^{\infty }dm\frac{-i%
\left[ 1+\left( -1\right) ^{2\left\vert m\right\vert }\right]
}{2\omega \left( n_{r}+\left\vert m\right\vert +\frac{1}{2}\right)
-V_{2}}\frac{\left( a_{+}^{\ast }\lambda _{+}\right)
^{n_{1}}\left( a_{-}^{\ast }\lambda _{-}\right) ^{n_{2}}}{\Gamma
\left( n_{1}+1\right) \Gamma \left( n_{2}+1\right) }.
\end{equation}
The poles of Green's function give the energy eigenvalues of the
system.  In order to obtain the Green's function of the physical
problem, the dummy coordinates should be eliminated. One of the
methods to eradicate a variable in the path integration formalism
is to integrate over it. The other method is to take the physical
eigenvalues of the corresponding conjugate momenta in the wave
function formalism. To eliminate the dummy coordinate $\phi $, we
prefer method in the wave function formalism \cite{Unal4}. \ We
take $m$\ as an azimuthal quantum number corresponding to the
operator $\widehat{p}_{\phi }$. Thus the Green's function is converted {\normalsize $\ $}into %
\begin{equation}
G^{\prime }\left( a_{\mp };\lambda _{\mp }\right)
=\sum\limits_{n_{r}=0}^{\infty }\left[ 1+\left( -1\right)
^{\left\vert m\right\vert }\right] \frac{\left( a_{+}^{\ast
}a_{-}^{\ast }\lambda
_{+}\lambda _{-}\right) ^{n_{r}+\left\vert m\right\vert }\left( \frac{%
a_{+}^{\ast }\lambda _{+}}{a_{-}^{\ast }\lambda _{-}}\right) ^{m}}{\sqrt{%
\Gamma \left( n_{1}+1\right) \Gamma \left( n_{2}+1\right) }}
\end{equation}
and physical coherent states become
\begin{equation}
\left\vert a_{\mp }\right\rangle =\sum\limits_{n_{r}=0}^{\infty
}\left[ 1+\left( -1\right) ^{\left\vert m\right\vert }\right]
\frac{\left(
a_{+}a_{-}\right) ^{n_{r}+\left\vert m\right\vert }\left( \frac{a_{+}}{a_{-}}%
\right) ^{m}}{\sqrt{\Gamma \left( n_{1}+1\right) \Gamma \left(
n_{2}+1\right) }}.
\end{equation}
As a function of physical coordinates $u$ and $t$, kernel of the
system is written as
\begin{equation}
K^{\prime }\left( u,\phi ;\lambda _{\mp }(\tau )\right) =\int
\frac{da_{\mp }^{\ast }da_{\mp }}{\left( 2\pi i\right)
^{2}}e^{-a_{\mp }^{\ast }a_{\mp }}\left\langle u,\phi \right\vert
\left. a_{\mp }\right\rangle K^{\prime }\left( a_{\mp }^{\ast
},\tau ;\lambda _{\mp },\tau _{a}\right) ,
\end{equation}
with $\left\langle u,\phi \right\vert \left. a_{\mp }\right\rangle
$ is
\begin{equation}
\left\langle u,\phi \right\vert \left. a_{\mp }\right\rangle =Ne^{-\frac{%
\left\vert a\right\vert ^{2}}{2}}\exp \left[ -\frac{1}{2}M\omega u^{2}+2%
\sqrt{M\omega }u\left( a_{+}e^{-i\phi }+a_{-}e^{i\phi }\right) -\frac{a^{2}}{%
2}\right].
\end{equation}
where $a^{2}=a_{+}a_{-}+a_{-}a_{+}$ and $N=\left( \frac{M\omega }{\pi }%
\right) ^{2}e^{-\frac{\left\vert a\right\vert ^{2}}{2}}$. If we
take integrations over $a_{\mp }^{\ast }$ , $a_{\mp }$ in Eq.
(31), it will be
\begin{equation*}
K\left( u,\phi ;\lambda _{\mp }\right) =N\exp \left[
-\frac{1}{2}M\omega u^{2}+2\sqrt{M\omega }ue^{-i\omega \tau
}\right.
\end{equation*}
\begin{equation}
\times \left. \left( \lambda _{+}e^{-i\phi }+\lambda _{-}e^{i\phi }\right) -%
\frac{\lambda ^{2}}{2}\right].
\end{equation}
The parametric time dependence of the eigenvalues of the operators
$\ \widehat{a_{j}}$ \ is the form of $a_{j}(\tau
)=a_{ja}e^{-i\omega \left( \tau -\tau _{a}\right) }$ . Let us
rewrite it in terms of trigonometric functions
\begin{equation}
K\left( u,\phi ,;\lambda _{\mp }(\tau )\right)
=Ne^{-\frac{1}{2}M\omega
u^{2}-\frac{\left\vert \lambda \right\vert ^{2}}{2}}\exp \left[ -2i\sqrt{%
\frac{M\omega u^{2}}{2}\left( i\lambda \right) ^{2}}\left(
e^{-i\left( \phi +\delta \right) }+e^{i\left( \phi +\delta \right)
}\right) \right].
\end{equation}
Here $\delta $ is complex phase and determined as
\begin{equation}
e^{-i\delta }=\frac{\lambda _{+}-i\lambda _{-}}{\sqrt{\lambda
_{+}\lambda +\lambda _{-}\lambda _{+}}}.
\end{equation}
To get coherent states in the parabolic coordinates, we express
the Kernel in terms of Bessel functions \cite{Gradshteyn}
\begin{equation*}
K\left( u,\phi ,;\lambda _{\mp }(\tau )\right)
=Ne^{-\frac{1}{2}M\omega u^{2}-\frac{\lambda
^{2}}{2}}\sum\limits_{m=-\infty }^{\infty }\left[ 1+\left(
-1\right)
^{\left\vert m\right\vert }\right] .
\end{equation*}
\begin{equation}
\times J_{m}\left( 2\sqrt{\frac{M\omega u^{2}}{2}\left( i\lambda \right) ^{2}%
}\right) e^{-im\left( \phi +\delta \right) }
\end{equation}
Integrating the Kernel over the parametric time $\tau $ Green's
function can be written
\begin{equation*}
G\left( u;\lambda _{\mp }\right) =\int \frac{dp_{0}}{2\pi
i}e^{-ip_{0}\left( t-t_{a}\right) }\sum\limits_{n_{r}=0}^{\infty
}\frac{i\left[ 1+\left( -1\right) ^{\left\vert m\right\vert
}\right] }{\omega \left[ n_{r}+\left\vert m\right\vert
/2+1/2\right] -V_{2}/2}~\Phi _{n_{r},m}(u)~
\end{equation*}%
\begin{equation}
\times \frac{e^{-\left\vert \lambda \right\vert ^{2}}e^{im\pi
/2}e^{-\lambda
^{2}/2}\left( -\lambda ^{2}\right) ^{n_{r}+\left\vert m\right\vert /2}}{%
\sqrt{\Gamma \left( n_{r}+\left\vert m\right\vert +1\right) }}\cos
m\left( \delta +\frac{\pi }{2}\right)
\end{equation}
where \bigskip $\Phi _{n_{r},m}(u)$ is
\begin{equation}
\Phi _{n_{r},m}(u)=\frac{e^{-\frac{1}{2}M\omega u^{2}}\left(
\sqrt{2}M\omega /\pi \right) ^{1/4}\left( \sqrt{\frac{M\omega
u^{2}}{2}}\right) ^{\left\vert
m\right\vert }}{\sqrt{\Gamma \left( n_{r}+\left\vert m\right\vert +1\right) }%
}L_{n_{r}+\left\vert m\right\vert }^{\left\vert m\right\vert }\left( \frac{%
M\omega u^{2}}{2}\right).
\end{equation}
Thus the Green's function of physical time in the parabolic
coordinates can be obtained as
\begin{equation*}
G(u,\phi ,t;\lambda _{\mp },t_{a})=\sum\limits_{n_{r}=0}^{\infty
}\sum\limits_{m=-\infty }^{\infty }e^{-iV_{2}\left( 1-\frac{\omega }{V_{2}}%
(n_{r}+1/2\right) ^{2}\left( t-t_{a}\right) }\frac{e^{im\phi
}}{2\pi }
\end{equation*}
\begin{equation*}
\times \frac{\left( \lambda _{+}\lambda _{-}\right)
^{n_{r}+\left\vert
m\right\vert }\left( \frac{\lambda _{+}}{\lambda _{-}}\right) ^{m}}{\sqrt{%
\Gamma \left( n_{1}+1\right) \Gamma \left( n_{2}+1\right) }}
\end{equation*}
\begin{equation}
\times \frac{\sqrt{MV_{2}/\left( n+1\right)
}e^{-\frac{1}{2}M\omega
u^{2}}\left( \sqrt{M\omega u^{2}/2}\right) ^{\left\vert m\right\vert /2}}{%
\pi ^{1/4}\sqrt{\Gamma \left( n_{r}+\left\vert m\right\vert +1\right) }}%
L_{n_{r}+\left\vert m\right\vert }^{\left\vert m\right\vert
}(\frac{M\omega u^{2}}{2}).
\end{equation}
where $n=n_{r}+\left\vert m\right\vert /2$. \ From Eq. (39) energy
eigenstates become
\begin{equation*}
\psi _{n_{r},m}(u)=\sqrt{\frac{2M}{n}}\frac{\sqrt{MV_{2}/2n+1}^{-\frac{1}{2}%
M\omega u^{2}}\left( \sqrt{M\omega u^{2}/2}\right) ^{\left\vert
m\right\vert
/2}}{\pi ^{1/4}\sqrt{\Gamma \left( n_{r}+\left\vert m\right\vert +1\right) }}%
L_{n_{r}+\left\vert m\right\vert }^{\left\vert m\right\vert }\left( \frac{%
M\omega u^{2}}{2}\right)
\end{equation*}
Using the residue of Green's function the energy eigenvalues of
the system becomes
\begin{equation}
E=-V_{2}\left[ 1-\frac{\omega }{V_{2}}(n_{r}+1/2)\right] ^{2}.
\end{equation}
The above result agrees with one available in literature~\cite{Sever2}.
\section{\protect\normalsize PT-symmetric and non-Hermitian Generalized
Morse case}
 If $V_{1}$ and$~V_{2}$ are real and $\alpha =i\alpha $ then The generalized Morse potential has the form
\begin{equation}
V(x)=V_{1}e^{-2i\alpha x}-V_{2}e^{-i\alpha x}.
\end{equation}
We can obtain the parametric time coherent states and energy
spectra for this potential following the steps of Sec. II by using
the same variables then the action becomes
\begin{equation*}
A=\int d\tau \left[ p_{u_{1}}\frac{du_{1}}{d\tau }+p_{u_{2}}\frac{du_{2}}{%
d\tau }-\omega \left[ \frac{p_{u_{1}}^{2}+p_{u_{2}}^{2}}{2M\omega }+\frac{1}{%
2}M\omega (u_{1}^{2}+u_{2}^{2})+\frac{V_{2}}{\omega }\right]
\right.
\end{equation*}
\begin{equation}
+\left. \left( -p_{0}\right) \frac{dt}{d\tau }+\frac{d\phi }{d\tau
}\left( p_{\phi }-\sqrt{2Mp_{0}}\right) \right]
\end{equation}
{\normalsize Frequency of the two oscillators system is again $\omega =\sqrt{%
\frac{-2V_{1}}{M}}$ and energy is $V_{2}$. Here we see that the
kinetic energy term is negative. By holomorphic coordinate
transformations the action becomes
\begin{equation}
A^{\prime }(a_{b}^{\dagger },\tau _{b};a_{a},\tau _{a})=\int d\tau
\left[ \frac{1}{2i}\left( \frac{da^{\dagger }}{d\tau }a-a^{\dagger
}\frac{da}{d\tau }\right) -\omega \left( a^{\dagger }a\right)
+V_{2}\right].
\end{equation}
The action of the system becomes as in Eq. (15) again. The form in
Eq. (16) takes
 \begin{equation}
K^{\prime }(a_{b}^{\dagger },\tau _{b};a_{a},\tau
_{a})=e^{i(-V_{2}-\omega)(\tau _{b}-\tau _{a})}\int \frac{Da^{\dagger}Da}{\left[ 2\pi i\right]^{2}}
\exp \left\{i\int\limits_{\tau _{a},a_{a}}^{\tau _{b},a_{b}^{\dagger }}d\tau \left[\frac{
1}{2i}\left( \frac{da^{\dagger }}{d\tau }a-a^{\dagger }\frac{da}{d\tau }
\right)-\omega \left( a^{\dagger }a\right)\right]\right\}.
\end{equation}
So, following the same steps of Sec. II, kernel of the system is
obtained as
\begin{equation}
K^{\prime }\left( u,\phi ,;\lambda _{\mp }(\tau )\right) =e^{-\frac{1}{2}%
M\omega u^{2}-\frac{\lambda _{+}\lambda _{-}}{2}}\exp \left[ -2i\sqrt{\frac{%
M\omega u^{2}}{2}\left( i\lambda \right) ^{2}}\left( e^{-i\left(
\phi +\delta \right) }+e^{i\left( \phi +\delta \right) }\right)
\right]
\end{equation}
and Green's function as a function of physical coordinates is written%
{\normalsize
\begin{equation}
G^{\prime }\left( u_{b},t_{b};\lambda _{\mp },t_{a}\right)
=\sum\limits_{n_{r}=0}^{\infty }\int\limits_{-\infty }^{\infty }dm\frac{i%
\left[ 1+\left( -1\right) ^{2\left\vert m\right\vert }\right] }{%
V_{2}/2-\omega \left( n_{r}+\left\vert m\right\vert +\frac{1}{2}\right) }%
\frac{\left( a_{+}^{\ast }\lambda _{+}\right) ^{n_{1}}\left(
a_{-}^{\ast }\lambda _{-}\right) ^{n_{2}}}{\Gamma \left(
n_{1}+1\right) \Gamma \left( n_{2}+1\right) }.
\end{equation}
Hence parametric time coherent states for PT-symmetric and
non-Hermitian Generalized Morse can be written as
\begin{equation}
\left\vert a_{\mp }\right\rangle =\sum\limits_{n_{r}=0}^{\infty
}\left[ 1+\left( -1\right) ^{\left\vert m\right\vert }\right]
\frac{\left(
a_{+}a_{-}\right) ^{n_{r}+\left\vert m\right\vert }\left( \frac{a_{+}}{a_{-}}%
\right) ^{m}}{\sqrt{\Gamma \left( n_{1}+1\right) \Gamma \left(
n_{2}+1\right). }}
\end{equation}
{\normalsize After performing integrations again, Green's function
of generalized Morse Potential takes}
\begin{equation*}
G\left( u;\lambda _{\mp }\right) =\int \frac{dp_{0}}{2\pi
i}e^{-ip_{0}\left( t-t_{a}\right) }\sum\limits_{n_{r}=0}^{\infty
}\frac{i\left[ 1+\left( -1\right) ^{\left\vert m\right\vert
}\right] }{V_{2}/2-\omega \left[ n_{r}+\left\vert m\right\vert
/2+1/2\right] }~\Phi _{n_{r},m}(u)~
\end{equation*}
\begin{equation}
\times \frac{e^{im\pi /2}e^{-\lambda ^{2}/2}\left( -\lambda
^{2}\right) ^{n_{r}+\left\vert m\right\vert /2}}{\sqrt{\Gamma
\left( n_{r}+\left\vert m\right\vert +1\right) }}\cos m\left(
\delta +\frac{\pi }{2}\right) ,
\end{equation}
where $\Phi _{n_{r},m}(u)$ is
\begin{equation}
\Phi _{n_{r},m}(u)=\frac{e^{-\frac{1}{2}M\omega u^{2}}\left( \sqrt{\frac{%
M\omega u^{2}}{2}}\right) ^{\left\vert m\right\vert
}}{\sqrt{\Gamma \left( n_{r}+\left\vert m\right\vert +1\right)
}}L_{n_{r}+\left\vert m\right\vert }^{\left\vert m\right\vert
}\left( \frac{M\omega u^{2}}{2}\right) .
\end{equation}
Thus for PT-symmetric and non-Hermitian generalized Morse
potential, Green's functions of physical time in the parabolic
coordinates can be obtained
\begin{equation*}
G(u,\phi ,t;\lambda _{\mp },t_{a})=\sum\limits_{n_{r}=0}^{\infty
}\sum\limits_{m=-\infty }^{\infty }e^{-iV_{2}\left( 1+\frac{\omega }{V_{2}}%
(n_{r}+1/2\right) ^{2}\left( t-t_{a}\right) }\frac{e^{im\phi
}}{2\pi }
\end{equation*}
\begin{equation*}
\times \frac{\left( \lambda _{+}\lambda _{-}\right)
^{n_{r}+\left\vert
m\right\vert }\left( \frac{\lambda _{+}}{\lambda _{-}}\right) ^{m}}{\sqrt{%
\Gamma \left( n_{1}+1\right) \Gamma \left( n_{2}+1\right) }}
\end{equation*}
\begin{equation}
\times \frac{\sqrt{MV_{2}/2n+1}e^{-\frac{1}{2}M\omega u^{2}}\left( \sqrt{%
M\omega u^{2}/2}\right) ^{\left\vert m\right\vert /2}}{\pi ^{1/4}\sqrt{%
\Gamma \left( n_{r}+\left\vert m\right\vert +1\right)
}}L_{n_{r}+\left\vert m\right\vert }^{\left\vert m\right\vert
}(\frac{M\omega u^{2}}{2}).
\end{equation}
The eigenstates of \ {\normalsize PT-symmetric and non-Hermitian
generalized Morse potential} are
\begin{equation}
 \psi _{n_{r},m}(u)=\sqrt{\frac{2M}{n}}\frac{\sqrt{MV_{2}\left( n+1\right) }%
e^{-\frac{1}{2}M\omega u^{2}}\left( \sqrt{M\omega u^{2}/2}\right)
^{\left\vert m\right\vert /2}}{\pi ^{1/4}\sqrt{\Gamma \left(
n_{r}+\left\vert m\right\vert +1\right) }}L_{n_{r}+\left\vert
m\right\vert }^{\left\vert m\right\vert }\left( \frac{M\omega
u^{2}}{2}\right).
\end{equation}
 {\normalsize Since the energy of the system is a residue of
Green's function, using the residue of Eq. $\left( 47\right)$,
we get}
\begin{equation}
E=-V_{2}\left[ 1+\frac{\omega }{V_{2}}(n+1/2)\right] ^{2}.
\end{equation}
From above equation, one can easily see that there are only real spectra for the PT-symmetric and non-hermitian Morse case~\cite{Berkdemir1,Berkdemir2,Sever1,Sever2,Yesiltas}.
{\normalsize \bigskip }
\section{\protect\normalsize Non-PT-symmetric and non-Hermitian generalized Morse case}
{\normalsize If we take $V_{1}=\left( A+iB\right)
^{2},~V_{2}=\left( 2C+1\right) \left( A+iB\right) $ and $\alpha
=1$ the generalized Morse potential can be rewritten as }
\begin{equation}
V(x)=\left( A+iB\right) ^{2}e^{-2x}-\left( 2C+1\right) \left(
A+iB\right) e^{-x}.
\end{equation}
Here $A,B,C$ are arbitrary parameters. This potential is non-PT
symmetric and non-Hermitian, but it has reel spectra. If $V_{1}$
is real $~V_{2}=A+iB$ and $\alpha =i\alpha $ is taken then the
Morse potential is transformed into the form
\begin{equation}
V(x)=V_{1}e^{-2i\alpha x}-\left( A+iB\right) e^{-i\alpha x}.
\end{equation}
Now we can derive coherent states and energy spectra following the
same steps in part II we get the action as
\begin{equation}
A=\int ds\left\{ \left( -p_{0}\right) \frac{dt}{d\tau }+\sqrt{2mp_{0}}\frac{%
d\phi }{d\tau }+A^{\prime }\right\}
\end{equation}
where $\phi $ is a dummy coordinate. Lagrange multiplier
$\sqrt{2mp_{o}}$ is added. Hence  the new action becomes
\begin{equation}
A^{\prime }=\int d\tau \left\{ p_{u_{1}}\frac{du_{1}}{d\tau }+p_{u_{2}}\frac{%
du_{2}}{d\tau }-\omega \left[ \frac{p_{u_{1}}^{2}+p_{u_{2}}^{2}}{2M\omega }+%
\frac{1}{2}M\omega \left( u_{1}^{2}+u_{2}^{2}\right) -\frac{A+iB}{\omega }%
\right] \right\}
\end{equation}
Here $A^{\prime }$ is two oscillators action and frequency $\omega =\sqrt{%
\frac{-V_{1}}{M}}$ and energy is $A+iB$. Here we notice that the
kinetic energy term is negative. The action can be written by
holomorphic coordinate transformation as
\begin{equation}
A^{\prime }(a_{b}^{\dagger },\tau _{b};a_{a},\tau _{a})=\int d\tau
\left[ \frac{1}{2i}\left( \frac{da^{\dagger }}{d\tau }a-a^{\dagger
}\frac{da}{d\tau }\right) -\omega \left( a^{\dagger }a\right)
-\left( A+iB\right) \right].
\end{equation}
The kernel of the system becomes
\begin{eqnarray}
K^{\prime }(a_{b}^{\dagger },\tau _{b};a_{a},\tau _{a})=\int \frac{%
Da^{\dagger }Da}{\left[ 2\pi i\right] ^{2}}\exp \left\{ i\int
d\tau \left[ \frac{1}{2i}\left( \frac{da^{\dagger }}{d\tau
}a-a^{\dagger }\frac{da}{d\tau }\right) \right. \right.\nonumber\\
\left. \left. -\omega \left( a^{\dagger }a\right) -\left(
A+iB\right) \right] \right\}.
\end{eqnarray}
So, Green's function of the physical system can be obtained as
follows
\begin{equation}
G^{\prime }\left( u_{b},t_{b};\lambda _{\mp },t_{a}\right)
=\sum\limits_{n_{r}=0}^{\infty }\int\limits_{-\infty }^{\infty }dm\frac{i%
\left[ 1+\left( -1\right) ^{2\left\vert m\right\vert }\right]
}{A+iB-\omega \left( n_{r}+\left\vert m\right\vert
+\frac{1}{2}\right) }\frac{\left( a_{+}^{\ast }\lambda _{+}\right)
^{n_{1}}\left( a_{-}^{\ast }\lambda _{-}\right) ^{n_{2}}}{\Gamma
\left( n_{1}+1\right) \Gamma \left( n_{2}+1\right) }
\end{equation}
and the parametric time coherent states for non-PT-symmetric and
non-Hermitian generalized Morse potential become are the same with
the ones given in Eq. (30)}. After performing integrations, wave
functions of the coherent states are
\begin{equation*}
G\left( u;\lambda _{\mp }\right) =\int \frac{dp_{0}}{2\pi
i}e^{-ip_{0}\left( t-t_{a}\right) }\sum\limits_{n_{r}=0}^{\infty
}\frac{i\left[ 1+\left( -1\right) ^{\left\vert m\right\vert
}\right] }{A+iB-\omega \left[ n_{r}+\left\vert m\right\vert
/2+1/2\right] }~\Phi _{n_{r},m}(u)~
\end{equation*}
\begin{equation}
\times \frac{e^{im\pi /2}e^{-\lambda ^{2}/2}\left( -\lambda
^{2}\right) ^{n_{r}+\left\vert m\right\vert /2}}{\sqrt{\Gamma
\left( n_{r}+\left\vert m\right\vert +1\right) }}\cos m\left(
\delta +\frac{\pi }{2}\right) ,
\end{equation}
where $\Phi _{n_{r},m}(u)$ is
\begin{equation}
\Phi _{n_{r},m}(u)=\frac{e^{-\frac{1}{2}M\omega u^{2}}\left( \sqrt{\frac{%
M\omega u^{2}}{2}}\right) ^{\left\vert m\right\vert
}}{\sqrt{\Gamma \left( n_{r}+\left\vert m\right\vert +1\right)
}}L_{n_{r}+\left\vert m\right\vert }^{\left\vert m\right\vert
}\left( \frac{M\omega u^{2}}{2}\right) .
\end{equation}
Therefore for non-PT-symmetric and non-Hermitian Morse potential,
the wave function of physical time in the parabolic coordinates
can be obtained as
\begin{equation*}
G(u,\phi ,t;\lambda _{\mp },t_{a})=\sum\limits_{n_{r}=0}^{\infty
}\sum\limits_{m=-\infty }^{\infty }e^{-i\left( A+iB\right) \left( 1+\frac{%
\omega }{A+iB}(n_{r}+1/2\right) ^{2}\left( t-t_{a}\right) }\frac{e^{im\phi }%
}{2\pi }
\end{equation*}
\begin{equation*}
\times \frac{\left( \lambda _{+}\lambda _{-}\right)
^{n_{r}+\left\vert
m\right\vert }\left( \frac{\lambda _{+}}{\lambda _{-}}\right) ^{m}}{\sqrt{%
\Gamma \left( n_{1}+1\right) \Gamma \left( n_{2}+1\right) }}
\end{equation*}
\begin{equation}
\times \frac{\sqrt{M\left( A+iB\right) \left( n+1\right) }e^{-\frac{1}{2}%
M\omega u^{2}}\left( \sqrt{M\omega u^{2}/2}\right) ^{\left\vert
m\right\vert
/2}}{\pi ^{1/4}\sqrt{\Gamma \left( n_{r}+\left\vert m\right\vert +1\right) }}%
L_{n_{r}+\left\vert m\right\vert }^{\left\vert m\right\vert
}(\frac{M\omega u^{2}}{2}).
\end{equation}
The eigenstates of non-PT-symmetric and non-Hermitian generalized
Morse case} are
\begin{equation}
\psi _{n_{r},m}(u)=\sqrt{\frac{2M}{n}}\frac{\sqrt{M\left( A+iB\right) n+1}%
e^{-\frac{1}{2}M\omega u^{2}}\left( \sqrt{M\omega u^{2}/2}\right)
^{\left\vert m\right\vert /2}}{\pi ^{1/4}\sqrt{\Gamma \left(
n_{r}+\left\vert m\right\vert +1\right) }}L_{n_{r}+\left\vert
m\right\vert }^{\left\vert m\right\vert }\left( \frac{M\omega
u^{2}}{2}\right).
\end{equation}
{\normalsize Since the energy of the system is a residue of Green's
function, from the residue of Eq. $\left(58\right)$, we obtain
energy eigenvalues as }
\begin{equation}
E=-(A+iB)\left[ 1+\frac{\omega }{A+iB}(n+1/2)\right] ^{2}.
\end{equation}
It is clear that energy spectra are real in cases $V_1>0$ if and only if $Re(V_2)=0$ and $V_1<0$ if and only if $Im(V_2)=0$.
\section{Conclusion}
We have studied parametric time coherent states for
PT-/non-PT-Symmetric and non-Hermitian generalized Morse potential
by using path integral formalism. Energy eigenvalues and the
corresponding wave functions are calculated. We have discussed the
negative energy coherent states. The wave functions of the
potential in Secs. III and IV are physical like Sec. II. Energy
eigenvalues are positive in contrary to expectation. Negative
energy coherent states can be obtained by analytic continuation
like \cite{Unal2}.
\section{Acknowledgements}
This research was partially supported by the Scientific and
Technological Research Council of Turkey.


\begin{thebibliography}{99}
\bibitem{Bender}
C. M. Bender and S.Boettcher, Pys. Lett., \textbf{80}, 5243
(1998); C. M. Bender, S. Boetcher, and P.N. Meisenger, J. Math.
Phys. \textbf{40}, 2201 (1999); C. M. Bender, G. V. Dunne, and P.
N. Meisenger, Phys. Lett. \textbf{A252}, 272 (1999).
\bibitem{Ahmed}
Z.Ahmed, Phys. Lett. \textbf{A282}, 343 (2001).
\bibitem{Bagchi}
B. Bagchi, C. Quesne, Phys. Lett. \textbf{A273}, 285 (2000).
\bibitem{Berkdemir2}
C. Berkdemir, A. Berkdemir and R.Sever, Physical Rev.
\textbf{C72},027001 (2005).
\bibitem{Znojil}
M. Znojil, Phys. Lett. \textbf{A264}, 108 (1999).
\bibitem{Feynman}
 R.P.Feynman, Hibbs, Quantum Mechanics and Path Integrals, McGraw-Hill, New York,
 (1965). chap. 1-2
\bibitem{Sever1}
N. Kandirmaz, R. Sever, Chin. J. Phys. \textbf{47}, 46 (1999)
[arXiv:0812.2614].
\bibitem{Morse}
P. M. Morse. Phys. Rev. \textbf {34}, 57 (1929).
\bibitem{Unal2}
N. Unal, Phys. Can. J. Phys. \textbf{80}, 875 (2002).
\bibitem{Berkdemir1}
 C. Berkdemir, J. Han, Chem. Phys.Lett. \textbf{409}, 203 (2005).
\bibitem{Duru1}
I.H. Duru, Phys. Rev. \textbf{D28}, 10 (1983).
\bibitem{Duru2}
I.H. Duru and H. Kleinert, Phys. Lett. \textbf{B84}, 185 (1979)
Fortschr. Phys. \textbf{30}, 401 (1982).
\bibitem{Sever2}
M. Aktas, R. Sever, Journal of Molecular Structure-Theochem
\textbf{710}, 223 (2004).
\bibitem{Schrodinger}
E. Schr\"{o}dinger, Naturwissenschaften, \textbf{14}, 664 (1926).
\bibitem{Unal1}
 N. Unal, Found Phys. \textbf{28}, 755 (1998).
\bibitem{Unal4}
 N. Kandirmaz, N. Unal, Theor. and Math. Phys. \textbf{155}, 884 (2008).
\bibitem{Gradshteyn}
I.S. Gradshteyn and I.M. Ryzhic:Table of Integrals, Series, and
Products,2nd ed. Academic Press, New York, (1981), chap.8.
\bibitem{Yesiltas}
O. Yesiltas, M. Simsek, R. Sever and C. Tezcan, Phys. Script. \textbf{67}, 472 (2003).


\end{thebibliography}
\end{document}